# "Ghost Cities" Analysis Based on Positioning Data in China

Guanghua Chi[a, b], Yu Liu[b], Zhengwei Wu[a], Haishan Wu[a*]

[a] Big Data Lab, Baidu Research, Baidu Inc., Beijing 100085, China
[b] Institute of Remote Sensing and Geographic Information Systems, Peking University, Beijing 100871, China

**Abstract**: Real estate projects are developed excessively in China in this decade. Many new housing districts are built, but they far exceed the actual demand in some cities. These cities with a high housing vacancy rate are called "ghost cities." The real situation of vacant housing areas in China has not been studied in previous research. This study, using Baidu positioning data, presents the spatial distribution of the vacant housing areas in China and classifies cities with a large vacant housing area as cities or tourism sites. To the best of our knowledge, it is the first time that we detected and analyzed the "ghost cities" in China at such fine scale. To understand the human dynamic in "ghost cities", we select one city and one tourism sites as cases to analyze the features of human dynamics. This study illustrates the capability of big data in sensing our cities objectively and comprehensively.

**Introduction**

China has experienced fast development during the past decade. From 1984 to 2010, the urban built-up area has increased from 8,842 km$^2$ to 41,768 km$^2$ (Nie & Liu, 2013). The urbanization speed is unprecedented in human history with so many buildings constructed in such a short time (Xue & Tsai, 2013). The amount of concrete used in China in the three years (2011-2013) is more than that used in U.S. in the 20$^{th}$ century (McCarthy, 2014). The fast urbanization of China has contributed to the high housing vacancy rate in some cities. Many new housing districts are built, but they far exceed the actual demand. In these cities, the population density is very low, and the residential districts are dark with few lights at night. Therefore, they are called "ghost cities." In Shepard's book "Ghost Cities of China" (Shepard, 2015), he defined "ghost city" as "a new development that is running at severe undercapacity, a place with drastically fewer people and businesses than there is available space for."

The "Ghost city" phenomenon has attracted much attention in recent years. Shepard (2015) said that "China is the world's most populated country without a doubt has the world's largest number of empty homes." Chinese Premier Keqiang Li warned the risk of rapid urbanization and said that "Urbanization is not about building big, sprawling cities. We should aim to avoid the typical urban malady where skyscrapers coexist with shanty towns" (Ryan, 2013). Zuoji Dong, head of the Ministry of Land and Resources planning bureau, said "new guidance issued by the ministry would allow for strict controls on new urban development. Unless a city's population is too dense or expansion is deemed necessary to cope with natural disasters, new urban districts will not be permitted" (Rafagopalan, 2014).

Media have reported many cities in China with a large vacant housing area. However, sometimes their opinions on a "ghost city" are completely opposite. For example, media have covered the serious situation of high vacant housing areas in Rushan City, while other media have reported that Rushan City has added residents and already expelled the term of "ghost." These reports, obtained by taking pictures or counting the number of homes with lights at night, have been criticized for their low credibility. Moreover, the internal causes of "ghost cities" may differ a lot despite the same outward appearance. Tourism in China has developed quickly recently. To satisfy the demand of tourists, cities with attractive tourism resources have built many houses for vacation. During the popular tourism seasons, many people will live in the vacant housing areas. While in other seasons, the population is small. Therefore, it is unfair to treat tourism sites and cities as equal. This raised the question: what is the real situation of "ghost cities" in China?

---

[*] Corresponding author.
E-mail address: guanghua@uw.edu (G. Chi), liuyu@urban.pku.edu.cn (Y. Liu),
wuzhengwei@baidu.com (Z. Wu), wuhaishan@baidu.com (H. Wu).

The vacant housing rate is one of the most important indicators to evaluate the health of real estate in each city. This indicator can also be used to discover the "ghost cities." The Chinese government has not published any data related to the vacant housing rate. The National Bureau of Statistics of China mentioned that the difficulty of calculating vacant housing rates lies in the difficulty to define a standard of the status of vacant houses and length of vacancy (House China, 2015). Besides using the vacant housing rate to define a ghost city, there are two other definitions. The Ministry of Housing and Urban-Rural Development of China gives a standard that 1 $km^2$ area holds 10 thousand people. Based on this standard, the rank of "ghost cities" in 2014 defines that cities with people smaller than half of the standard are "ghost cities" (Su, 2014). The first three rank cities based on this indicator are Erenhot, Qinzhou and Lahsa. Chen (2014) proposed an alternative equation to calculate the indicator of "ghost cities": $(S-D)/n$, where $S$ is the supply of new houses in the following five years. $D$ is the demand of new houses in the following five years. $n$ is the number of houses at present. This equation reflects the proportion of current houses that should be removed to satisfy the balance between supply and demand. The first three rank cities based on this indicator are Ordos, Yingkou and Ulanqab.

It is difficult to obtain real estate data and population data with high spatial resolution. This impedes the understanding of "ghost cities." Su and Chen's studies cannot pinpoint the exact location of vacant housing areas and can only reflect the average level of "ghost" in each city, not to mention finding out the reasons behind "ghost cities." These results would be questionable since they are aggregated results. Fortunately, the emergence of big data brings opportunities to objectively understand the status of or even reasons behind "ghost cities." Widely-applied location-aware devices (LAD), such as mobile phones and GPS receivers, generate large volumes of individual trajectory data with long time scale and high resolution. These features make it suitable for population analysis. For example, Kang et al. (2012) used mobile phone data to estimate the population distribution in China. It provides a mean for observing urban dynamics from a micro perspective, including the human migration and interaction between regions. According to the social sensing concept proposed by Liu et al. (2015), we can use the data generated by each individual to sense our living environment.

Whether the so called "ghost cities" have a high housing vacancy has always been disputed for the lack of data to verify. In this study, we use the location data of mobile APPs and point of interest data of residential area of Baidu, the largest search engine company in China. The basic idea of discovering vacant housing areas is that only a small population lives in these areas. To the best of our knowledge, it is the first time that we detected and analyzed the "ghost cities" in China at such fine scale. We should note that we do not attempt to give the rank of which areas have the most "ghost cities." Instead, we want to find out the exact location of vacant housing areas at the moment. Even if the vacant housing area is very large in a city, it does not mean that this city will still have a high vacant ratio in the future. Like the Zhengdong new district, it has been developing quickly in the past few years and attracted a large number of people.

**Methodology**

Data Description

This study uses two types of datasets, including Baidu positioning data and points of interests (POI). The attributes of Baidu positioning data cover anonymized user id, latitude, longitude, and time. There are several billions of positioning points each day. The time span is from 9/8/2014 to 4/22/2015. Its features of national spatial scale, long temporal scale, and high precision make our study of "ghost cities" representative and reliable. We admit that this dataset is biased. It cannot represent all the demography of a city, such as very young and very old people, or those who do not use smart mobile phones. However, it can represent the situation of the population density, which is the focus of our study. The POI data include POI name, latitude, longitude, and category. The specific processing of POI data would be discussed in the next section.

Discovering and Classifying Vacant Housing Area

The basic idea of discovering vacant housing areas is that only a small population lives in a residential area. Therefore, we should calculate two variables based on our data: users' home location and location of residential areas. We first adopted DBSCAN algorithm to calculate the home location of each user based on their positioning points from 9:00 am to 6:00 pm. DBSCAN is a widely used density based clustering algorithm and is more computationally efficient than other methods like MeanShift (Ester et al., 1996). It stands for density-based spatial clustering of applications with noise. It is based on the idea of density reachability: many points that are closely located will be grouped together as clusters, while points that are located in low-density regions are considered as noise. This algorithm requires two parameters: the minimum distance $\varepsilon$ to decide whether two points are reachable and the minimum number of points *minPts* to form a cluster. A point $P_1$ is reachable to point $P_n$ if there are a sequence of points $P_1, P_2, ..., P_n$ where each $P_i$ is within $\varepsilon$ to $P_{i-1}$. After experimenting with different values of these two parameters, we set $\varepsilon$ as 200 meters and *minPts* as 2. We choose the center of the cluster with largest number of points as the user's location. Cases exist where users move their houses during the period of our data. But they only cover a low proportion and will not affect our results. We keep both locations of these users in our study.

Our study attempts to discover the spatial distribution of vacant housing areas. Therefore, it is crucial to know exactly the location of residential areas. We use POI data from Baidu Map, which has high reliability and quality. We select POIs whose categories are residential areas and villas. We delete the residential area POIs that are within one kilometer of villas since population density near villas is very low. Moreover, it is possible that the residential area is built recently and few people live there. We will remove these situations based on the population living there and the detail will be discussed blow.

The difficulties of discovering vacant housing areas based on positioning points and POI lie in two aspects. First, the residential area POI is a point, while residential district is an area. The residential area POI is not always located in the center of the residential area. Second, the areas of residential districts are not the same. In other words, they have different length and width. If we have the data of residential area polygons instead of points, we can directly count the exact population in each residential area. Such data may be available in some residential areas of few big cities. But the vacant housing area is usually small in these big cities. While for those cities with a large vacant housing area, we do not have such data. To solve this problem, we use 100 m * 100 m grids as the analysis units in this study. As shown in Figure 1, for each residential area POI (Point A), we select 5 * 5 grids with the residential area POI in the center grid. We select the top six most populated grids. If the sum is less than 300, we define this residential area as the vacant housing area. The reason why we choose the threshold as 300 will be explained blow. We also set the sum more than 60 to exclude those recently built residential areas where few people live.

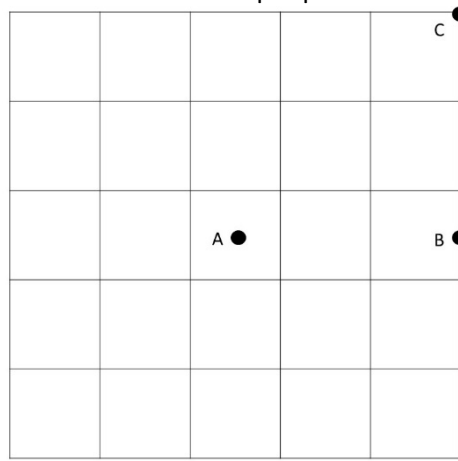

Fig 1. Illustration of the study area based on residential area POIs (This figure shows the study area generated based on residential area POI A. Residential area points B and C will be discussed blow. Each grid is 100m * 100m. )

The edge length of a residential area is between 300 meter and 500 meter on average. The location of a residential area POI is not always located in the center of a residential area. It may be located at the edge of a residential area (Point B in Fig. 1). At the extreme situation, it may be located at the vertex of a residential area (Point C in Fig. 1). In this extreme case, only a fourth part of the 500 m * 500 m grids would cover the residential area. The other three fourths part of the area may be covered with non-residential area. If we summarize all the population in the 500 m * 500 m grids, results in cases like point B and C in Figure 1 would be distorted. This is the reason why we only select the first six grids with the most population.

The average floor area ratio is 1 (An, 2015) and the average living area is 30 $m^2$ in China (Ruan, 2010). Thus a 100 m * 100 m grid can hold 333 people. The dataset we used in this study contains 770 million users. China has a population of 1.36 billion. Thus on average a 100 m * 100 m grid can hold 188 Baidu users. We define that a grid with people smaller than one fourth of the standard is a vacant housing area. If the six grids with most population holds people smaller than 300, this area is defined as the vacant housing area.

We should note that the first six grids with most population may be located in the nearby residential area of the study residential area rather than itself. We admit that our results fail to discover some vacant housing areas. However, the vacant housing areas are normally located spatially aggregated. Even though one residential area is lost, its nearby residential areas will be discovered. Therefore, this special case will not cause too many noises to the final result.

To classify those tourism sites with a large vacant housing area, we use the positioning data in China's National Day (9/29/2014 and 10/2/2014), New Year's Day (12/30/2014 and 1/2/2015), and International Workers' Day (4/29/2015 and 5/2/2015) in the vacant housing areas. By comparing the change of population before and after the holiday, we can judge whether the vacant housing area is for tourism. If the population in a vacant housing area increases at the holiday, it is defined as a tourism area.

**Results**

We count the total number of all vacant housing areas POI in each county. We also count the area of all vacant housing areas and find that the result is similar. We select 20 cities with a large vacant housing area. The result is shown in Figure 2, which can be browsed interactively on the website (http://bdl.baidu.com/ghostcity/). This study presents the spatial distribution of vacant housing areas. We should note that the rank of these 20 cities is not the real rank of our result, but they all rank at the top of 50. The reason why we did not show the real rank is that the rank is very sensitive and it may affect the sale of real estate.

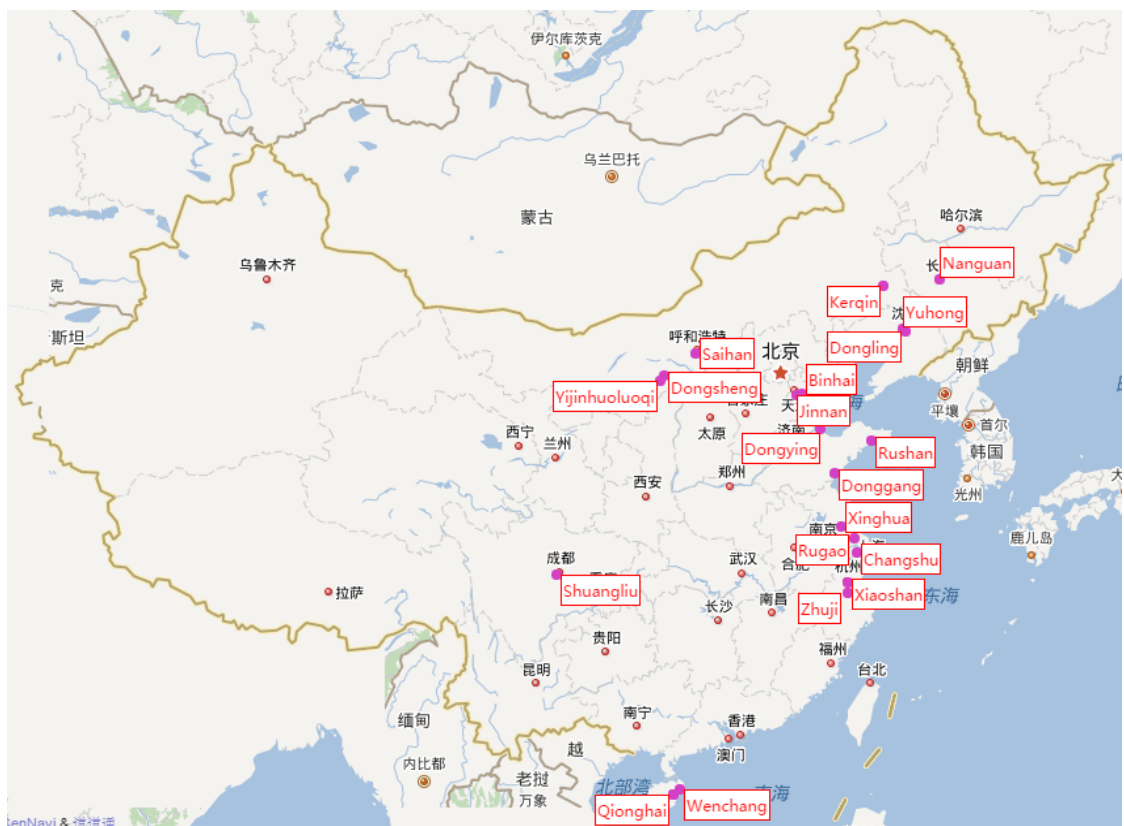

Fig. 2 20 cities with a large vacant housing area

From a macro view, cities with a large vacant housing area are mostly second-tier and third-tier cities. The tiers of cities in China are determined by their income, education, transportation, technology and other indicators. East provinces have more proportion of cities with vacant housing areas. From the micro view, vacant housing areas are mostly distributed in the city's periphery or new towns of cities. As shown in Figure 3A, the 500 m * 500 m grid covers the residential area. Even though it does not cover the other part of the residential area, the result has already shown that the residential area is a vacant housing area. Figure 3B shows two residential areas of vacant housing area. The location of the left-bottom residential area in Figure 3B is similar with the point C in Figure 1. Our algorithm exactly finds this area.

We present the spatial distribution of the first 8 cities plus Dongling district (Liaoning Province) with a large vacant housing area in Figure 4. All these 9 cities have been reported by media for their large vacant housing area. In figure 4A, the vacant housing areas distribute tightly and are close to the sea in Rushan. Media has reported that Rushan built too many see-view houses for vacation. The vacant housing rate is high except in summer. Figure 4B and C show the spatial distribution of vacant housing area in Ordos, which is well known as "ghost city." The vacant housing areas mostly are located in the city's periphery. Figure 4H is located near the figure 4C and has the similar situation of vacant housing rate. Figure 4D is Binhai New Area, which was called "the third increasing point of economy in China" ever. However, this new area exists many half-baked buildings. For Rugao, Dongying, and Xinghua (Fig. 4E, F, G), we can find the support from the Internet or media that these cities contain a large vacant housing area. Media reported that Shenyang National Games Village has become "ghost town" (Miller & Chow, 2015). Our result confirmed this report and showed the real spatial distribution of vacant housing areas (Fig. 4I).

We also detect whether these cities are tourism sites. We find that many tourism sites are located near the sea, such as Rushan (Shandong Province), Rugao (Jiangsu Province), Xinghua (Jiangsu Province), Wenchang (Hainan Province), Donggang (Shandong province), and Qionghai (Hainan Province). The other two cities like Zhuji (Zhejiang Province) and Xiaoshan (Zhejiang Province) also have popular tourism resource.

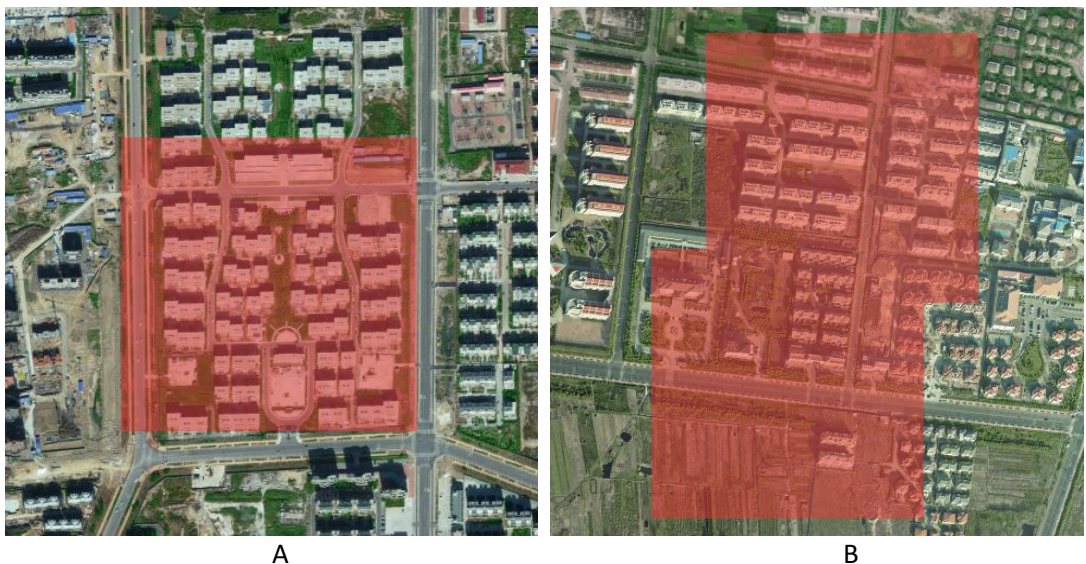

A                                          B

Fig. 3 Illustration of residential areas of vacant housing area we discovered. (A. One residential area B. Two residential areas)

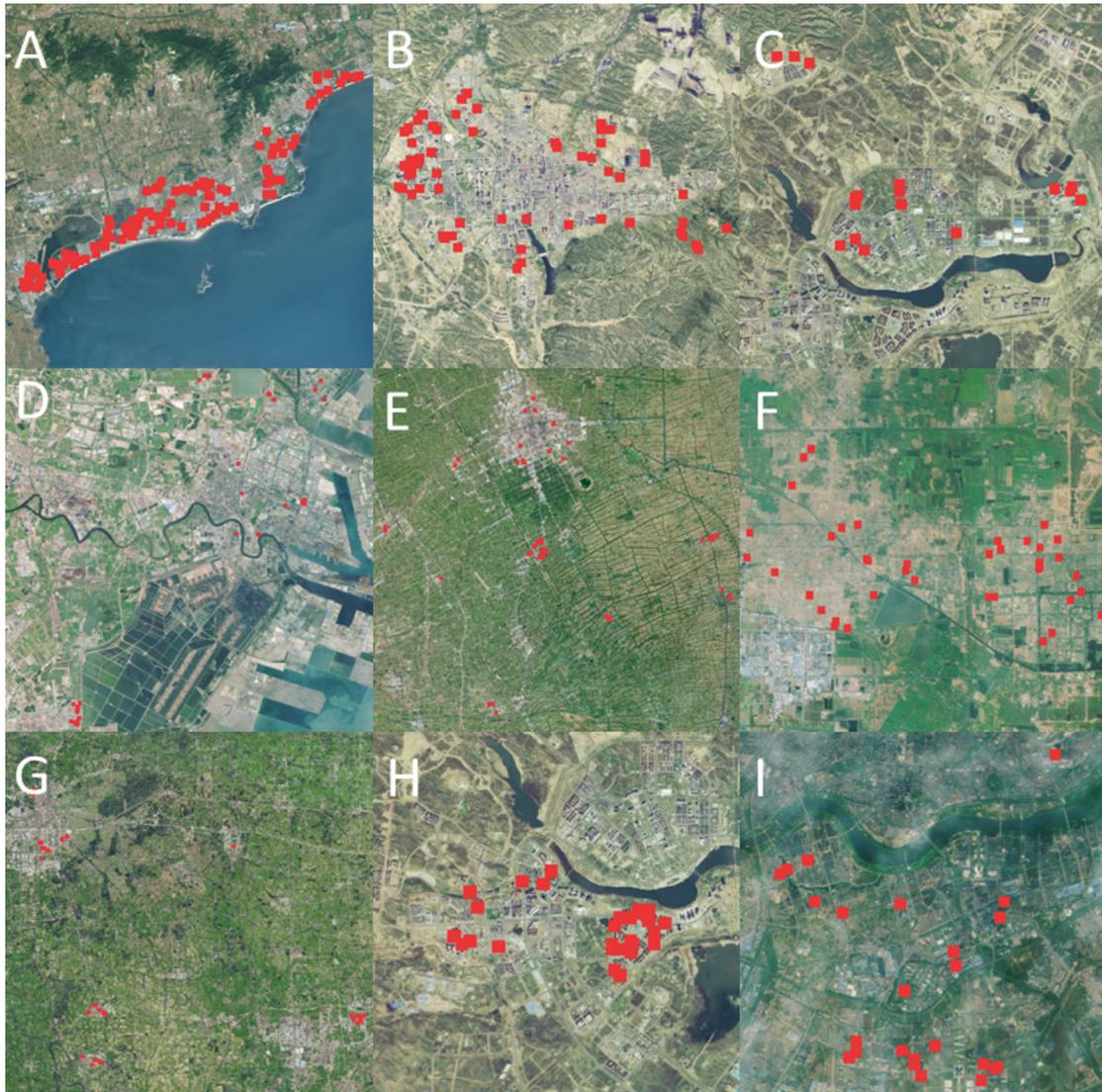

Fig. 4 The vacant housing areas in 9 cities. (Regions overlaid by red grids are inferred as vacant housing areas. A. Rushan B. Dongsheng C. Ordos D. Binhai E. Rugao F. Dongying G. Xinghua H. Yijinhuoluoqi I. Dongling)

**Case Study**

To find out the real situation of cities with a large vacant housing area and discover the reason why they are called "ghost cities", we select Rushan (a tourist site) and Kangbashi (a city) as cases to analyze the population change, home-work separation, and human migration. These two cities are well-known for their large vacant housing area. Kangbashi is a new district, which once belonged to Dongsheng. The Ordos government moved from Dongsheng to Kangbashi in 2006 to increase its development speed. Kangbashi contains abundant coal and other natural resources, which boosts its economic development quickly. As it became richer, the government ambitiously starts building a new city. Much capital was invested in residential real estate, which became an investment instead of living demand. The housing vacancy ratio is very high. For more information, please refer to (Woodworth, 2015).

Rushan is located in Shandong province. It has a 21 kilometer coastline with a beautiful sea view, which is called Rushan Yintan, meaning silver beach. Most of the residential real estate are seasonal houses. Many people bought a house there for vacation. Every summer a large number of tourists will visit there, while after the summer the population will decrease dramatically. For these two different types of cities with a large vacant housing area, we attempt to find out the features of human dynamic.

Population Dynamics

We count the population in these two cities each day and show the result in Figure 5. Because the number of Baidu users is increasing, we should remove this effect to show the real population change in each city. In this study, we use the population in each city divided by the total number of people in China as the indicator of population. Then we normalize the value of these two cities. Kangbashi has a clear weekly cycle of population change, while Rushan is not. This is related to their functional difference. Rushan is a tourist site and people there do not have a work related cycle. In National Day, Rushan has the largest population and the number is much more than other days in whole year. The population in Kangbashi decreases, while in Rushan the number increase. In China's Spring New Year, Kangbashi has the fewest number of people. The population in Rushan is fewest at the beginning of February instead of Chinese Spring New Year. In New Year, the population in Rushan does not increase significantly because the weather is cold and few people visit there for vacation. In Qingming Festival (2015/4/5), the population in Kangbashi decreases significantly, while the number in Rushan increases slightly. Rushan is a tourist site while Kangbashi is not, which leads to the difference of population change in Rushan and Kangbashi. From September in 2014 to April in 2015, the number of people in these two large vacant housing area did not increase, showing no trend of attracting more people. However, we should note that Rome wasn't built in a day; neither are new cities in China. Constructing a new city is easy, while making it functional needs a long-term endeavor (Shepard, 2015).

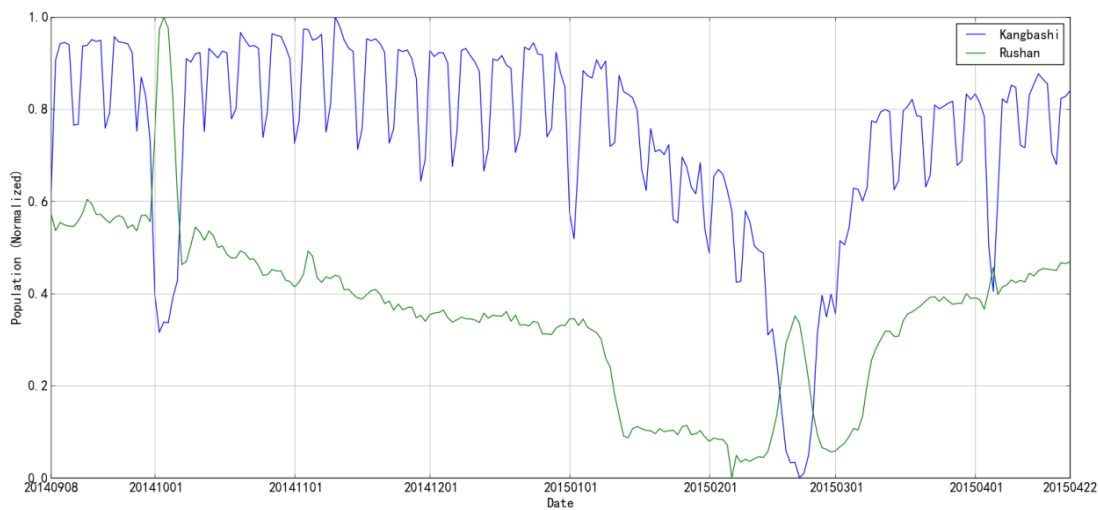

Fig. 5 Population change of Kangbashi and Rushan

Figure 6 shows the population change in each hour of one week from 2015/3/23 to 2015/3/29. Similar to the results from mobile phone data (Kang et al., 2012), the largest population appears at 12:00pm and 20:00pm. The two peaks are almost similar in Kangbashi, while in Rushan the peak in 20:00pm is higher. In weekends, the population in Kangbashi decreases, while the number in Rushan keeps stable.

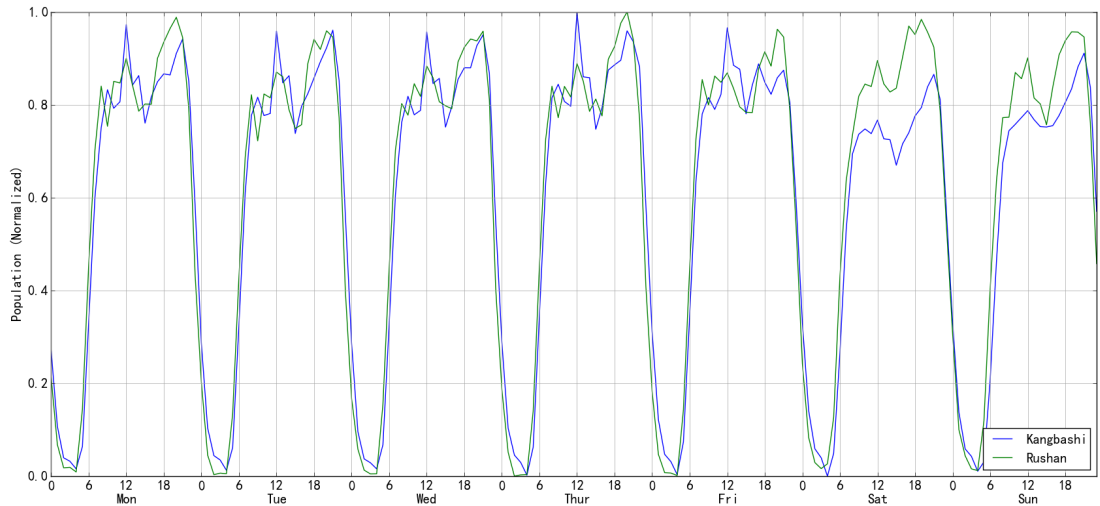

Fig. 6 Population change in each hour of a week

Home-work Analysis

Home-work separation is a normal phenomenon in modern city life, which means that a considerable distance exists between a person's home and work place (Poston, 1972). This study attempts to understand the home-work separation situation in "ghost cities." The results can help government learn about the real situation and make reasonable planning. We used DBSCAN algorithm to calculate the work place based on a person's positioning points from 9:00am to 6:00pm.

We adopted the algorithm designed by Eric Fischer (Fischer, 2013) to visualize the spatial distribution of people's home and work places. This algorithm assigns lightness value to each pixel based on the number of points in the pixel. As shown in Figure 7a, most of work places are located closely to Kangbashi government seats. In Rushan Yintan, the number of home and work places is really small compared with Rushan downtown since it is a tourist area (Fig. 7b). Rushan government realized the situation of few work opportunities in Rushan Yintan and began to attract companies to increase the employment.

We statistic the proportion of people who live or work in Kangbashi and Rushan Yintan (Table 1). It is a common sense that the proportion of people who work and live in the same city contains most part (Shi et al., 2015). Administrative boundaries effectively impact the human migration across the territory (Chi et al., 2014). The number of people who live in Dongsheng and work in Kangbashi is twice as many as that who live in Kangbashi and work in Dongsheng. Dongsheng has a perfect infrastructure. People prefer to live in Dongsheng and work in Kangbashi. This illustrates that Kangbashi should increase more job opportunities and improve its infrastructure. A new city being beautiful and modern cannot attract people migrate there. Jobs, industry, entertainment, health and education systems should be functioning before many people migrate to a new city (Shepard, 2015). These two cases reflect that vacant housing areas usually contain limited job opportunities, and basic infrastructure should be improved to satisfy people's need.

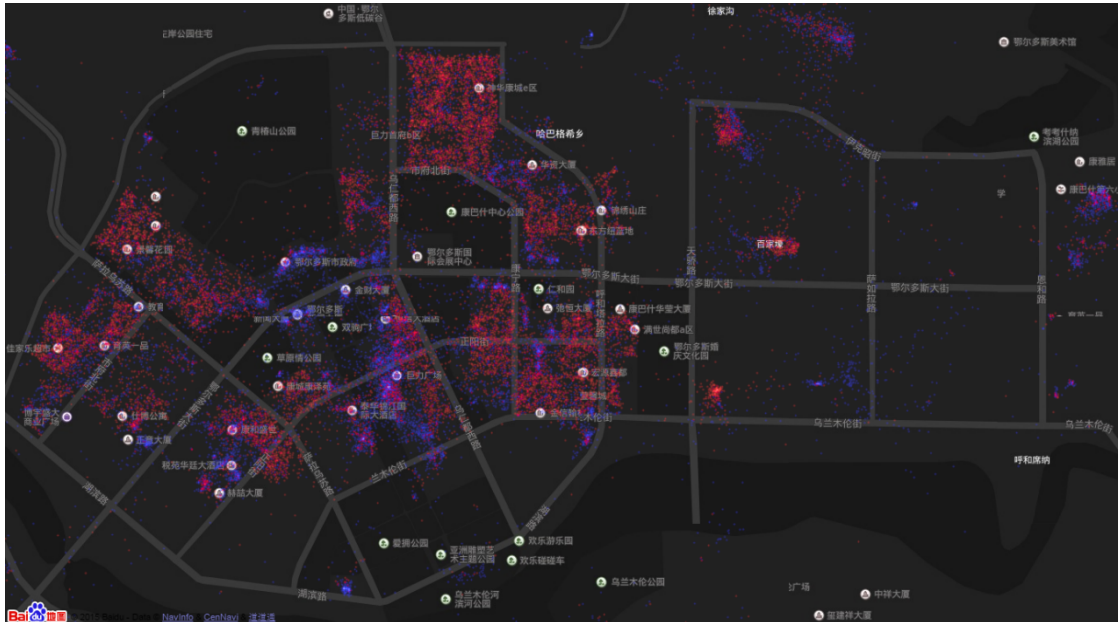

a. Kangbashi

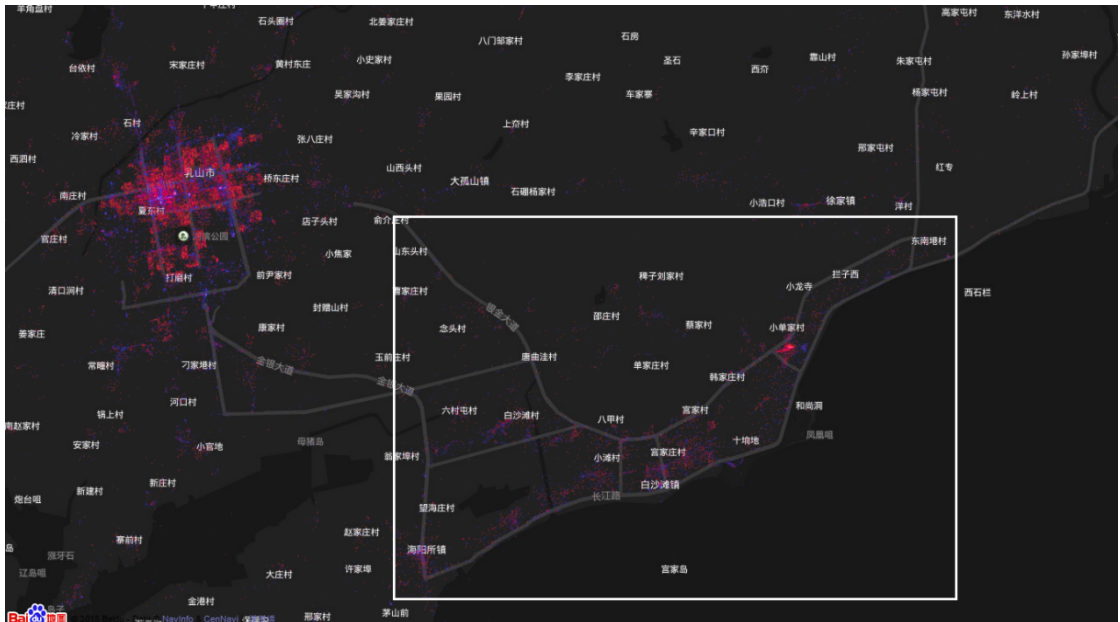

b. Rushan (The white polygon is Rushan Yintan where sea view houses are located.)
Fig. 7 The spatial distribution of home and work place (Red points denote home. Blue points are work places.)

Table. 1 Home and work place distribution in Rushan and Kangbashi

| Kangbashi | | | Rushan | | |
| --- | --- | --- | --- | --- | --- |
| Home | Work place | Proportion（%） | Home | Work place | Proportion（%） |
| Kangbashi | Kangbashi | 24.9 | Rushan Yintan | Rushan Yintan | 33.5 |
| Kangbashi | Dongsheng | 8.2 | Rushan Yintan | Rushan downtown | 8.1 |

| Kangbashi | Other | 24.7 | Rushan Yintan | Other | 18.3 |
| Dongsheng | Kangbashi | 17.6 | Rushan downtown | Rushan Yintan | 16.6 |
| Other | Kangbashi | 24.5 | Other | Rushan Yintan | 23.5 |

## Human Migration

Human migration reflects the urban vitality. The tempo-spatial trajectory is composed of a person's positioning points ordered by time. In this section, we attempt to discover the city interaction caused by human migration. If a user moves from city *a* to city *b*, we suppose that city *a* and *b* has an interaction. When calculating human migration, it is unreasonable to define that a user has a migration between two cities as long as he/she exposed in two different cities because he/she may pass by cities between the origin city and the destination city in space. For example, as shown in table 2, a person moves from Guangzhou to Shanghai at Nov. 1$^{st}$, 2014. He/she passed by Changsha, Nanchang, and Hangzhou before arriving at Shanghai. It is wrong to suppose that Guangzhou and Changshan has an interaction. In calculating migration, the difficult part is how to remove pass-by cities. The frequency of positioning points exposed by people is not stable. The number of positioning points exposed in pass-by cities may be larger than that exposed in destinations. These situations make the calculation of human migration more difficult. Meanwhile, the large amount of our dataset requires that the algorithm has a high efficiency.

This study supposes that the migration will last for more than one day. We ignore the cases that people return back in one day. We define a person's trajectory as $\{t_1, city_1\}$,  $\{t_2, city_2\}$, $\{t_3, city_3\}$, ...,  $\{t_n, city_n\}$. $t_n$ is the time of the *n*th positioning point. $city_n$ is the city of the *n*th positioning point. We first select the city of the first positioning point each day as people's location. If $city_n$ is same to $city_{n-1}$, then we merge these two records. Otherwise, we suppose that the person migrates from $city_{n-1}$ to $city_n$.

Table 2 A sample of a person's trajectory

| **Province** | **City** | **Time** |
|---|---|---|
| Guangdong | Guangzhou | 2014/11/1 10:30:01 |
| Hunan | Changsha | 2014/11/1 12:50:01 |
| Jiangxi | Nanchang | 2014/11/1 14:52:01 |
| Zhejiang | Hangzhou | 2014/11/1 18:30:01 |
| Shanghai | Shanghai | 2014/11/1 19:31:01 |
| Shanghai | Shanghai | 2014/11/2 09:21:01 |
| Shanghai | Shanghai | 2014/11/3 08:13:01 |

Figure 8 shows the result of human migration in these two cities. It makes sense that the number of people inflow or outflow increases during holidays. The migration population increases as the number of Baidu users increases. Compared to Kangbashi in National Day, a peak of population inflow appears firstly in Rushan, then followed by a peak of population outflow. While in Kangbashi, the peak of population inflow and outflow appears almost at the same time. In these two "ghost cities", we did not discover the increase of net population flow, meaning that they did not attract more people live there. Besides the change of migration population, we also know the interaction intensity between two cities. In other words, we know the proportion of migration people from/to each city. Figure 9 shows the location and proportion of migration population in National Day. The source and sink with largest flow in Kangbashi and Rushan are Yulin and Weihai, respectively.

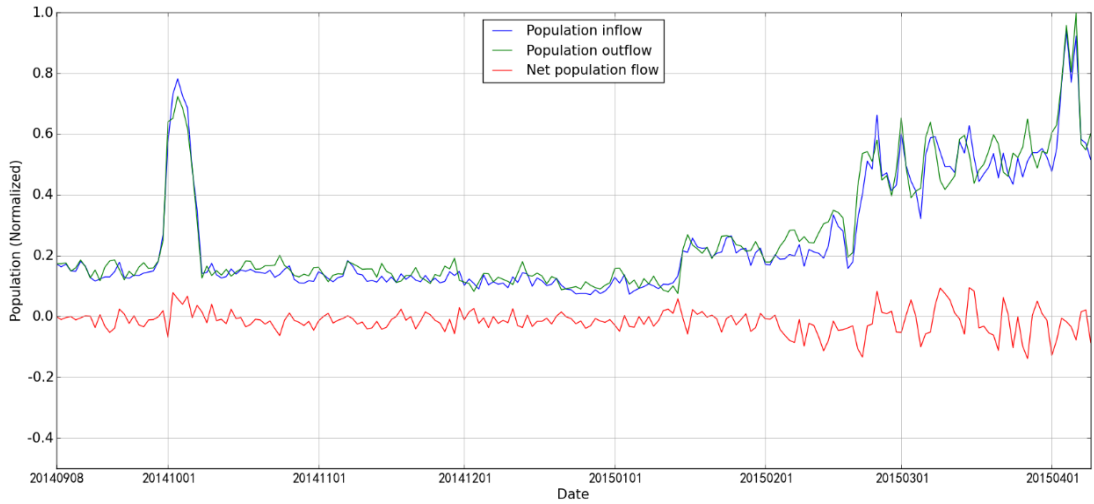

a. Kangbashi

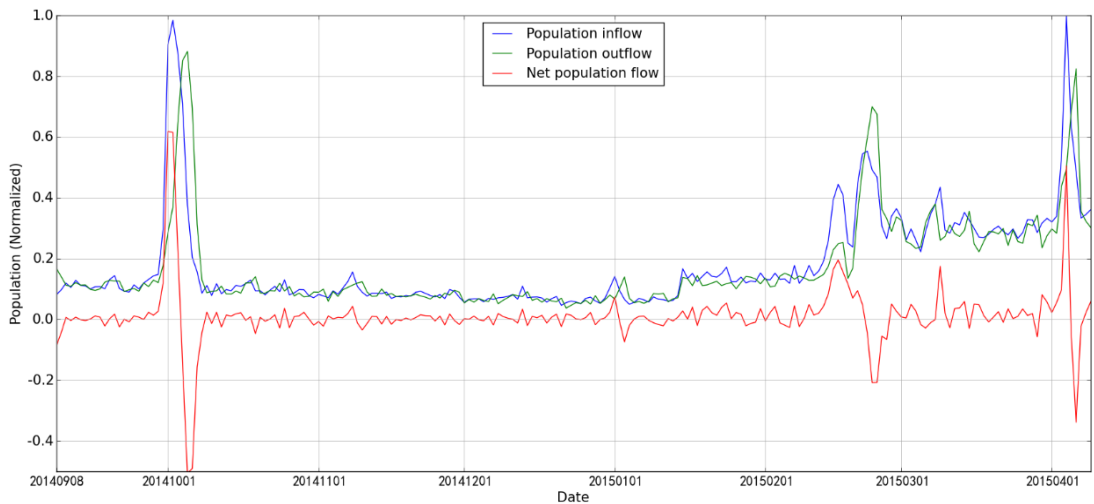

b. Rushan

Fig. 8 Population change of migration

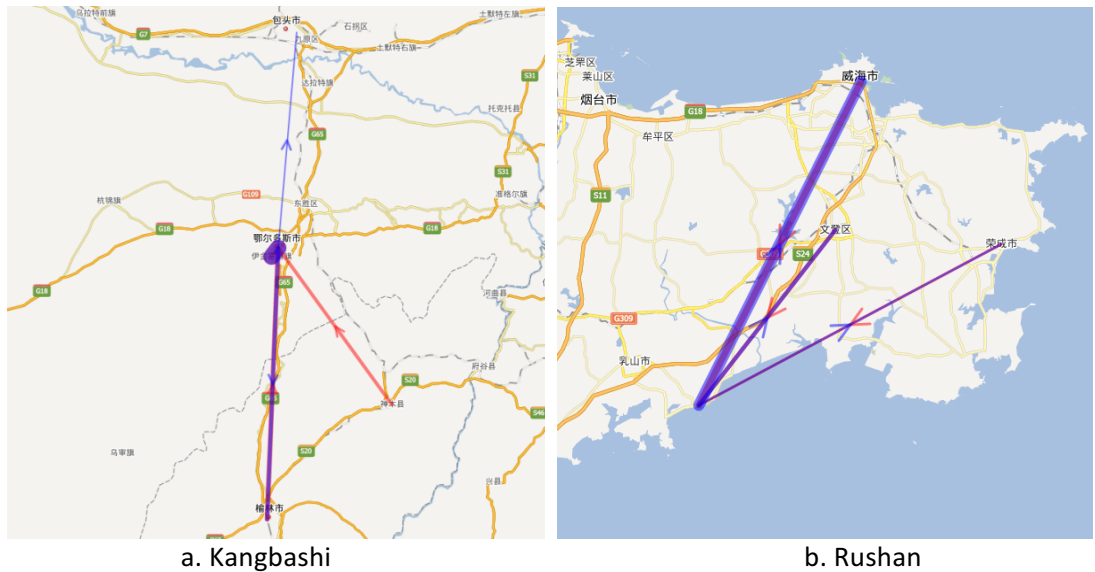

|   a. Kangbashi   |   b. Rushan   |

Fig. 9 Human migration in National Day

**Conclusion**

The fast urbanization of China has contributed so many housing built that has outstripped the actual demand. Whether the so called "ghost cities" reported by media have a high housing vacancy has always been disputed for the lack of data to verify. As far as we are concerned, we use big data to analyze the real situation of "ghost cities" in China for the first time. The features of national spatial scale, long temporal scale, and high precision of Baidu big data make the study of "ghost cities" representative and reliable. Instead of just counting the number of homes with light at night in certain residential areas as the indicator of "ghost city", Baidu big data can count the population precisely, in real time, and in national scale. A limitation of the data is that it cannot represent the real demography of a city because not all people are Baidu users. However, with the ubiquity of smart mobile phones, Baidu users occupy the most proportion of the whole population. Moreover, the quality of residential area POIs will affect our results. We make a series of processing to make sure that the POIs are reliable. Baidu big data bring opportunities to objectively understand the status or even reasons of "ghost cities."

Based on the Baidu positioning data and residential area POI data, we design an algorithm to discover the vacant housing areas. The results discovered the specific location of vacant housing areas, which can help government make smarter and more reasonable decisions. Our results provide the real situation of the so called "ghost cities" in China. Cities with a large vacant housing area are mostly second-tier and third-tier cities. The tiers of cities in China are determined by their income, education, transportation, technology and other indicators. East provinces have more proportion of cities with vacant housing areas. We also distinguish the tourism sites and cities. Based on Baidu positioning data, we discover the human dynamic in cities with a large vacant housing area to help better understand the situation in "ghost cities."

**Acknowledgements**

The authors would like to thank their colleagues in Big Data Lab Zhengxue Li for his data visualization in web design, Wei Jia for his help in pre-processing the data, and Geography Librarian Amanda Hornby at University of Washington for her insightful comments.